\documentclass[useAMS]{mn2e}
\usepackage{epsfig}
\title[Raman Scattered He~II in Symbiotic Stars]
{On the Center Shifts of Raman Scattered He~II and H$\alpha$ Wings 
in Symbiotic Stars}
\author[Jung and Lee]{Yang Chan Jung \thanks{E-mail:
ycjung@arcsec.sejong.ac.kr} and Hee-Won Lee 
\thanks{E-mail: hwlee@sejong.ac.kr}\\
Department of Astronomy and Space Sciences \\ 
Astrophysical Research Center for the Structure and Evolution
of the Cosmos \\ 
Sejong University, Seoul, 143-747, Korea }

\begin{document}

\date{Accepted 1988 December 15. Received 1988 December 14; in original form 1988 October 11}

\pagerange{\pageref{firstpage}--\pageref{lastpage}} \pubyear{2004}

\maketitle

\label{firstpage}

\begin{abstract}
Using a Monte Carlo technique, we investigate the center shifts 
that are expected to occur for the broad H$\alpha$ wings and the 
He~II~$\lambda$~6545 feature as a function of the neutral hydrogen column 
density in symbiotic stars. 
These two features are proposed to be formed via Raman scattering
of UV continuum around Ly$\beta$ and He~II~$\lambda$~1025 emission
line by neutral hydrogen. The strengths of these two features
are determined by a combination of various physical parameters 
including the covering
factor and the neutral hydrogen column density $(N_{HI})$ of 
the scattering region. 
The branching ratio of Raman scattering to Rayleigh scattering
for UV radiation around Ly$\beta$ is a non-linearly increasing function of
the wavelength, which results in enhanced Raman optical fluxes redward 
of the H$\alpha$ line center as $N_{HI}$
increases.  However, we find that the amount of wing center shift is quite
small about $20{\rm\ km\ s^{-1}}$ as $N_{HI}$ increases 
from $10^{20}{\rm\ cm^{-2}}$ to $10^{21}{\rm\ cm^{-2}}$. 
Assuming that He~II~$\lambda$~1025 emission
is characterized by a Gaussian profile, the Raman scattered
He~II~6545 feature exhibits near Gaussian profiles with the peak
shifted redward for $N_{HI}<10^{22}{\rm\ cm^{-2}}$.
The redward center shift amounts to $1{\rm\ \AA}$ for $N_{HI}\sim
10^{20}{\rm\ cm^{-2}}$ and decreases as $N_{HI}$
increases up to $N_{HI}\sim 10^{22}{\rm\ cm^{-2}}$, above which no center
shift is observed.  The redward peak shift is due to 
the fact that the incident emission profile is symmetric with respect to 
He~II~$\lambda$~1025 line center whereas the Raman conversion rate 
is increasing towards the line center of Ly$\beta$. 
We emphasize that the determination of $N_{HI}$ by locating the exact peak 
position of the He~II~6545 feature will lift the degeneracy to allow a more 
accurate estimate of the covering factor of the neutral region, 
providing strong constraints on the mass loss process 
occurring in symbiotic stars.

\end{abstract}

\begin{keywords}
scattering --- radiative transfer --- binaries: symbiotic 
--- mass-loss --- circumstellar matter 
\end{keywords}

\section{Introduction}

Symbiotic stars are generally believed to be binary systems of a mass
losing giant and a hot star, which is usually a white dwarf (Kenyon
1986). The giant component loses a large amount of mass 
in the form of a slow stellar wind, a significant
fraction of which may be ionized by the strong UV radiation emanating
from the hot component (e.g. Taylor \& Seaquist 1984). 
The physical extent of the ionized region
should be sensitively dependent on the strength of the ionizing
radiation and the mass loss rate of the giant component. 
In this respect, studies of ionization structures of symbiotic stars 
are very important in understanding the mass loss process 
in symbiotic stars (e.g. Skopal 2003).

Symbiotic stars exhibit very unique spectroscopic features, which are
Raman scattered lines by neutral hydrogen. The emission features
around 6825 \AA\ and 7082 \AA\ appearing in more than half of the
symbiotic stars are Raman scattered O~VI 1032, 1038 doublets first
identified by Schmid (1989). The Raman scattering origin for these features
is strongly supported by many spectropolarimetric observations 
(e.g. Schmid \& Schild 1994, Schmid et al. 2000, Harries \& Howarth 1996) 
and contemporaneous far UV observations (e.g. Espey et al. 1995, 
Birriel, Espey, Schulte-Ladbeck 1998, 2000).  
The Raman scattering cross sections for O~VI~$\lambda\lambda$ 1032,1038 
are of order $10^{-22}{\rm\ cm^{2}}$. Therefore, the existence
of these features implies that a non-negligible fraction of the
scattering region is characterized by a very high H~I column densities
$N_{HI}\ge 10^{22}{\rm\ cm^{-2}}$.

Van Groningen (1993) found other Raman scattered
features including Raman scattered He~II lines 
in the symbiotic star RR Telescopii. In particular, Raman
scattered He~II line at around 4850 \AA\ was found in the spectrum 
of the young planetary nebula NGC 7027 by P\'equignot(1997).  
Because a He~II ion has the same atomic structure with four times 
large level spacings as a hydrogen atom does, the
emission lines associated with the energy levels with even principal
quantum numbers of He~II have a slightly shorter wavelengths 
than those of H~I
emission lines.  This implies that the Raman scattering cross sections
for these He~II lines are so high that the scattering processes can
operate in broader range of objects including young planetary
nebulae (Nussbaumer, Schmid \& Vogel 1989). 

In the same line of reasoning, UV continuum radiation around
Ly$\beta$ will be Raman scattered to be redistributed around H$\alpha$,
forming broad wings. In this case, the wing profiles reflect the
wavelength dependence of the scattering cross section, which is
approximately inversely proportional to the squared 
wavelength difference of the incident radiation and 
the Ly$\beta$ line center.  
These broad H$\alpha$ wings with profiles well-fitted 
by a profile $\propto\Delta\lambda^{-2}$
were found in objects including many symbiotic stars (Lee 2000), 
some young planetary nebulae such as 
IC 4997 (Lee \& Hyung 2000) and M2-9 (Balick 1989),
and post AGB stars (van de Steene, Wood, \& van Hoof 2000). 
Recently Arrieta \& Torres-Peimbert (2003) added 12 more 
objects that are also in late stages of stellar evolution.
Considerable mass loss is a common feature for all of these objects, where
one can find a large amount of neutral matter surrounding the central 
hot stars.  It is also interesting that a large fraction of these
objects exhibit bipolar nebular morphology (e.g. Schwarz \& Monteiro 2003,
Schwarz \& Corradi 1992).

Lee et al. (2003) used the 3.6~m Canada-France-Hawaii telescope to
perform spectroscopy of the symbiotic star V1016~Cygni. 
They found the Raman scattered He~II 6545 that is
converted from He~II~$\lambda$~1025 emission line. They adopted a simple
scattering geometry to compute the broad H$\alpha$ wing template
profiles and isolate the 6545 feature, from which 
they were able to put a lower limit of the mass loss rate of the
Mira variable. However, the strengths of Raman scattered features
are determined by a combination of the covering factor and
neutral hydrogen column density ($N_{HI}$) of the scattering region 
for given incident UV radiation. The degeneracy of the two factors 
should be lifted in order to put more stringent constraints 
on the mass loss processes occurring in symbiotic stars.

There exists conspicuous asymmetry in the branching ratio 
of Raman scattering relative to Rayleigh scattering 
around Ly$\alpha$, where the branching ratio
is an increasing function of the incident wavelength (Yoo, Bak, \& Lee 2002). 
Therefore, continuum radiation redward of Ly$\beta$ will be more 
easily Raman scattered than continuum blueward of Ly$\beta$. 
This tendency increases for higher
scattering optical depths, which may lead to a redward shift of
the center location of H$\alpha$ wings as $N_{HI}$ increases. Furthermore,
the Raman conversion rate increases in the wavelength region, where
He~II~$\lambda$~1025 is located. Therefore, if He~II~$\lambda$~1025 exhibits
a symmetric emission profile with the peak at its line center, 
the peak position of the Raman scattered He~II~6545 will move
toward H$\alpha$, where the amount of shift will be mainly determined by
the neutral hydrogen column density.

In this paper, we perform Monte Carlo simulations to compute
the center shifts of the H$\alpha$ wings and the Raman scattered He~II~6545
features for various $N_{HI}$, which may be useful
in lifting the degeneracy of the column density and the covering factor
that determine the strengths of the Raman scattered features. 
In the following section we briefly describe the basic ideas of 
the center shifts.
The numerical results from our Monte Carlo calculations are presented in
Section 3. Finally in Section 4 we discuss our results and observational
ramifications.

\section{Models and Monte Carlo Procedures} 

In a symbiotic star, the gravity of the white dwarf significantly
affects the stellar wind flow from the giant, as is shown 
by the SPH computations by Mastrodemos \& Morris (1998). 
However, it will still hold that the
matter density will decrease as the distance from the giant center
increases. As a first approximation, the ionization
structure was investigated by Taylor \& Seaquist (1984), assuming that
the circumstellar distribution is radial. With this assumption, the
ionized region will take a conical shape, which is determined by
the relative strength of H ionizing UV radiation $L_{H}$ 
and the mass loss rate $\dot M$ of the giant.  They introduced a parameter
$X_{H^{+}}$ defined as
\begin{equation}
X_{H^{+}} \equiv {4\pi \mu^2 m_p^2\over a_{B}(H,T_{e})a^{2}_{H}} D L_{H}
\left({v_{\infty} \over \dot M}\right)^{-2}
\end{equation}

Here, $a_{B}(H,T_{e})$ is the recombination coefficient of hydrogen in a 
region with electron temperature $T_{e},D$ is the binary separation, 
$\mu$ is the mean molecular weight and $a_{H}$ is the relative 
hydrogen particle density. 
The Str\"omgren boundary is formed where the parameter $X_{H^{+}}$ exceeds
the value of the function $f(\theta)$ defined by
\begin{equation}
f(\theta) \equiv {1\over \sin^3\theta} 
\left({\pi-\theta \over 2} +{\sin 2\theta \over 4}
\right),
\end{equation}
where the angle $\theta$ is the opening angle of the ionized conical 
region (Seaquist, Taylor \& Button 1984).

The Raman scattering optical depth of unity for UV
radiation corresponding to this spectral region is obtained when the
neutral hydrogen column density exceeds $N_{HI}\sim 10^{20}{\rm\ cm^{-2}}$
(e.g. Nussbaumer et al. 1989, Lee 2000). Therefore,
the Raman scattered He~II 6545 and nearby H$\alpha$ wings are formed
in the neutral region, because the neutral column density of 
the ionized region
does not exceed the column density $N_{HI}=10^{18}{\rm\ cm^{-2}}$, for
which the Lyman limit radiation becomes optically thick.
The neutral column density $N_{HI}$ for a given line of sight starting
from the UV emission region through the neutral scattering region 
will be inversely proportional to the impact parameter with respect to
the giant. Therefore, the determination of $N_{HI}$ may put a strong
constraint on the mass loss rate of the giant component 
in a symbiotic star.

In this paper, we are mainly concerned with
the Raman scattered He~II 6545 feature and broad H$\alpha$ wings found
in symbiotic stars.
We adopt a view that the broad H$\alpha$ wings are formed through
Raman scattering of continuum around Ly$\beta$, as is proposed by
Nussbaumer et al. (1989). The relative strength of Raman scattered
He~II~6545 feature with respect to that of H$\alpha$ wings are
determined basically by the equivalent width of He~II$\lambda$~1025
and the scattering optical depth. 

\begin{figure}
\epsfig{file=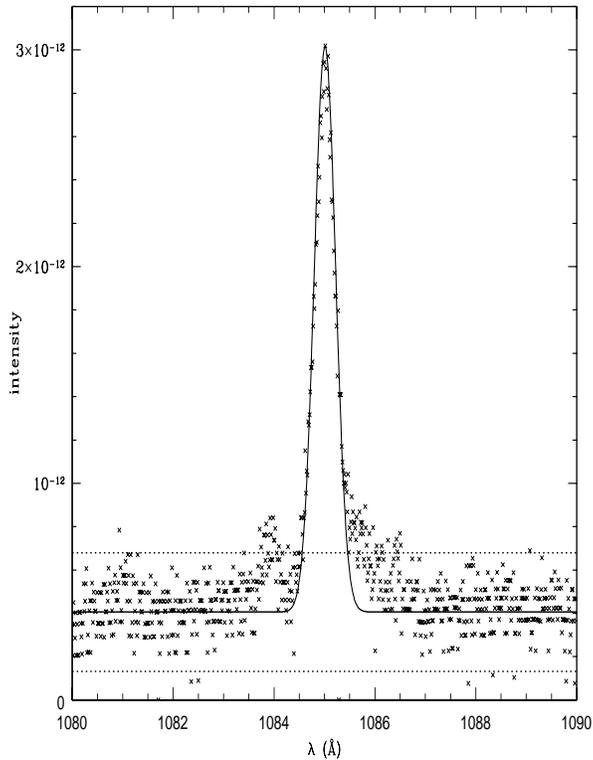, height=11cm, width=8cm}
 \caption{A spectrum around He~II~$\lambda$~1085 of the symbiotic
star RR Telescopii obtained with the ORFEUS.
The He~II~$\lambda$~1085 line is well fitted by a single Gaussian
$f_\lambda\propto f_0 e^{-(\lambda-\lambda_0)^2/\Delta\lambda^2}$, where
$\Delta\lambda =0.1 {\rm\ \AA}$ and $f_0 = 4.0\times 10^{-12}
{\rm\ erg\ s^{-1}\ cm^{-2}\ \AA^{-1}}$. The continuum is shown by a
solid line and dotted lines represent $2.5\sigma$, where the
statistics were obtained from the wavelength interval.
}
\end{figure}
Schmid et al. (1999) studied O~VI~$\lambda\lambda$~1032, 1038 lines
using ORFEUS (Orbiting and Retrievable
Far and Extreme Ultraviolet Spectrometers). 
In Fig.~1, we show a spectrum around He~II~$\lambda$~1085 of the symbiotic
star RR Telescopii obtained with the ORFEUS, in order to obtain a hint
of He~II~$\lambda$~1025 which suffers heavy interstellar extinction 
due to closeness to H~I Ly$\beta$. It is well fitted by a single Gaussian
$f_\lambda\propto f_0 e^{-(\lambda-\lambda_0)^2/\Delta\lambda^2}$, where
$\Delta\lambda =0.1 {\rm\ \AA}$ and $f_0 = 4.0\times 10^{-12}
{\rm\ erg\ s^{-1}\ cm^{-2}\ \AA^{-1}}$. 
In order to estimate the equivalent width of He~II~$\lambda$~1085,
we fit the continuum and 
the He~II~$\lambda$~1025 emission line by a solid line in the figure, 
from which the continuum flux density is chosen to be 
$f_c = 4\times 10^{-13}{\rm\ erg\ cm^{-2}\ s^{-1}\ \AA^{-1}}$.
By dotted lines we show the continuum levels higher and lower by
$2.5\sigma$.  We obtain the equivalent width $3.7{\rm\ \AA}$, 
and put a lower bound $2.0{\rm\ \AA}$ from this measurement. 
Using the Case B recombination
theory we may infer the equivalent width of He~$\lambda$~1025 with
the additional assumption that the continuum level does not change
much. In the current work, we adopt a value $0.5{\rm\ \AA}$, which
appears to fit to the spectrum of V1016~Cyg 
obtained by Lee et al. (2003). 

When the scattering optical depth is small, the Raman conversion
efficiency will be approximately proportional to the cross section that
is again approximately given by $\propto (\lambda-\lambda_{Ly\beta})^{-2}$
(e.g. Lee \& Lee 1997, Lee \& Hyung 2000).
Therefore, the peak position of the Raman feature is determined
by the location of the maximum of the function
\begin{equation}
p(\lambda)={p_0 e^{-[(\lambda-\lambda_{He^+})/
\Delta\lambda]^2}\over (\lambda-\lambda_{Ly\beta})^{2}},
\end{equation}
where $\lambda_{He^+}$ and $\lambda_{Ly\beta}$ are
the center wavelengths of He~II~$\lambda$~1025 and Ly$\beta$,
respectively. Here, $\Delta\lambda=0.08{\rm\ \AA}$ is the width of
the single Gaussian for incident He~II~$\lambda$~1025 emission line, which
was chosen in this work. Here, we define $Y\equiv \lambda_{Ly\beta}
-\lambda_{He^+}$, and $m_e$ and $m_p$ are masses of an electron and a proton, 
respectively.  In this work, we choose $\lambda_{Ly\beta}=1025.7222{\rm\ \AA}$
and $\lambda_{He^+}=1025.273{\rm\ \AA}$.
In Fig.~2, we show the Raman scattering cross section and the He~II~1025
emission line given by a single Gaussian with $\Delta\lambda
=0.08{\rm\ \AA}$. 
\begin{figure}
\epsfig{file=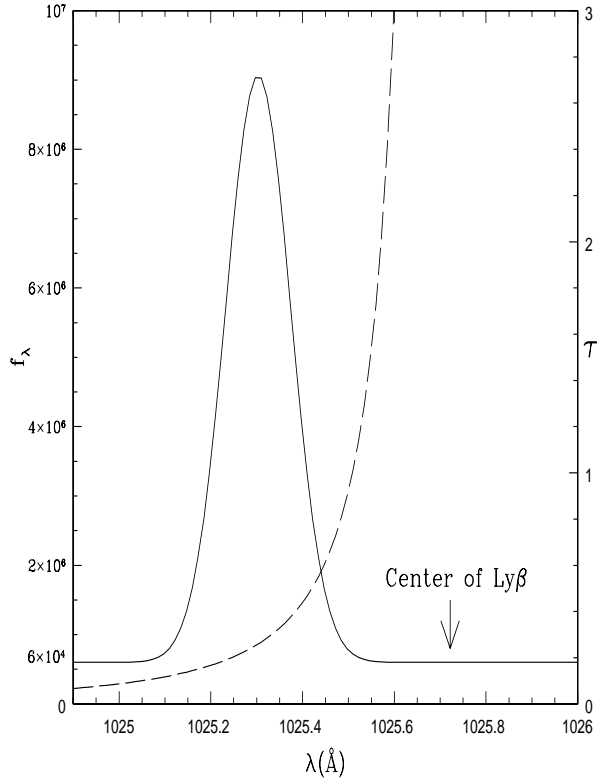, height=11cm, width=8cm}
 \caption{
The Raman scattering cross section overplotted to the He~II~1025
emission line given by a single Gaussian with $\Delta\lambda
=0.08{\rm\ \AA}$. Note that the cross section increases steeply toward
the Ly$\beta$ line center, which will affect the center location
of the Raman scattered He~II~6545 feature.
}
\end{figure}

By performing a simple differentiation, we find 
\begin{equation}
\lambda_{max,i} = \lambda_{He^+}+{Y-\sqrt{Y^2-4\Delta\lambda^2} \over 2},
\end{equation}
where $\lambda_{max, i}$ is the wavelength around $\lambda_{He^+}$,
at which the Raman scattered He~II will exhibit the peak.  In the limit 
$\Delta\lambda \ll Y$, the peak of Raman
conversion rate is found at
\begin{equation}
\lambda_{max,i} \simeq \lambda_{He^+}+\Delta\lambda^2/Y
\simeq \lambda_{He^+}+0.016{\rm\ \AA},
\end{equation}
for our assumed value $\Delta\lambda=0.08{\rm\ \AA}$. It is
interesting to note that the amount of peak shift is proportional to
the squared width of the incident radiation with a Gaussian profile.
This shift is translated into the redward shift of Raman
scattered He~II~6545 feature by an amount
\begin{equation}
\lambda_{max,o} \simeq \lambda_{He^+,Ram} +0.016\times 6.4^2{\rm\ \AA}=
\lambda_{He^+,Ram} +0.66{\rm\ \AA},
\end{equation}
which is quite significant and should be easily observable.

As the scattering optical depth increases, a significant fraction of
Raman scattered photons are formed after a few Rayleigh scatterings,
resulting in a constant Raman conversion rate, under the assumption of the
absence of absorbing opacities such as dust and little opacity near H$\alpha$
(e.g. Schmid 1996, Lee \& Lee 1997).  In this highly
optically thick limit, no shift in peak position will be seen.
This implies that the redward shift of Raman scattered feature
He~II~6545 decreases as the neutral hydrogen column density increases,
rendering the amount of redward shift a useful indicator of
$N_{HI}$. The multiple scattering effects
are difficult to handle analytically, and we adopt a Monte Carlo approach
to obtain the quantitative relation between the peak shift and 
$N_{HI}$.

Monte Carlo techniques have been applied usefully to understand basic
properties of Rayleigh and Raman scattered radiation and interpret various
spectroscopic and polarimetric observational data (e.g. Schmid 1995, 1996).
In this work, we use the Monte Carlo code developed by 
Yoo, Bak \& Lee (2002), in which they adopted the branching ratio $r_b$ of 
Raman  and Rayleigh scattering cross sections around Ly$\beta$ 
given by
\begin{equation}
r_b = 0.1342 + 12.50 \left(\lambda-\lambda_{Ly\beta} \over \lambda_{Ly\beta}
\right).
\end{equation}
This situation is illustrated in Fig.~1 of Yoo et al. (2002), where
the branching ratio was represented by a dotted line.  
Although it is highly unrealistic to characterize the scattering region 
by a single value of $N_{HI}$, we adopt the scattering regions in this way 
in order to find the amount of center shift dependent on $N_{HI}$.

The Raman scattering branching ratio increases as the wavelength 
increases fairly steeply according to Yoo et al. (2002).
For higher column densities the Raman wings will extend 
to broader range of wavelengths, where the difference in the branching ratios 
redward and blueward of the line center will increase significantly,
strengthening the red wing part selectively. 
This effect may lead to the redward shift of the H$\alpha$ wing center 
as $N_{HI}$ increases. 
 
\section{Results}
\subsection{Center Shift of H$\alpha$ Wings}
\begin{figure}
\epsfig{file=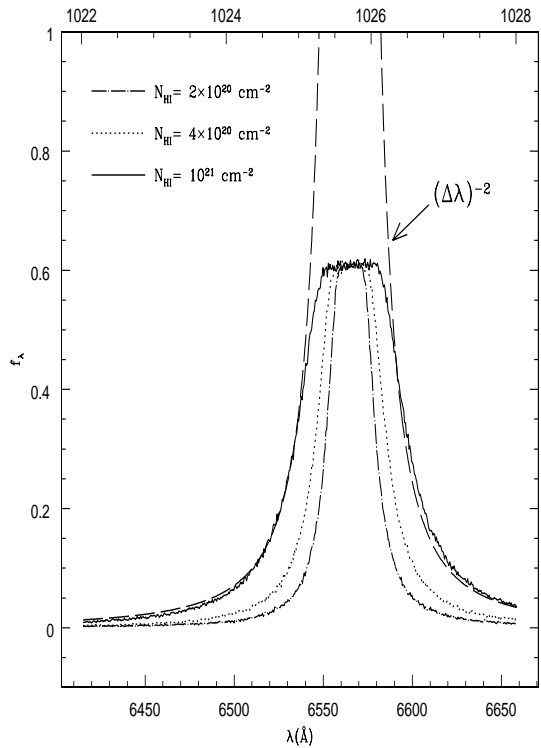, height=11cm, width=8cm}
\caption{
H$\alpha$ wings produced by our Monte Carlo code for 
Raman scattering of continuum radiation around Ly$\beta$ for various 
neutral hydrogen column densities in order to determine the wing center. 
The vertical axis represents the Raman conversion rate $f_\lambda$.
The dashed line shows the best fitting profile with the functional
dependence $\propto\Delta \lambda^{-2}=(\lambda-\lambda_0)^{-2}$
in the range $f_\lambda <0.5$.
}
\end{figure}
In Fig.~3, we show H$\alpha$ wings produced by our Monte Carlo code for 
Raman scattering of continuum radiation around Ly$\beta$ for various 
neutral hydrogen column densities in order to determine the wing center. 
The vertical axis represents the Raman conversion rate, which is the
ratio of the number of incident UV photons to that converted into optical
ones. As is explained by Lee \& Hyung (2000), when the medium becomes optically
thick, the conversion rate saturates to a constant value near 0.6. 

A least square method was adopted to locate the wing center. Using the
fact that the far wing parts are fairly well approximated by 
profiles $\propto\Delta \lambda^{-2}=(\lambda-\lambda_0)^{-2}$, 
by varying $\lambda_0$ we obtained a $\Delta
\lambda^{-2}$ profile that best fits each Monte Carlo result for those parts
with the conversion rate less than 0.5. For illustrative purpose, 
one such fit for $N_{HI}=
10^{21}{\rm\ cm^{-2}}$ is shown in Fig.~3 by a long-dashed line.
Even though $(\lambda-\lambda_{H\alpha})^{-2}$ profiles provide 
overall satisfactory fits to simulated H$\alpha$ profiles, it should be
noted that small deviations are also apparent especially near H$\alpha$
center.

\begin{figure}
\epsfig{file=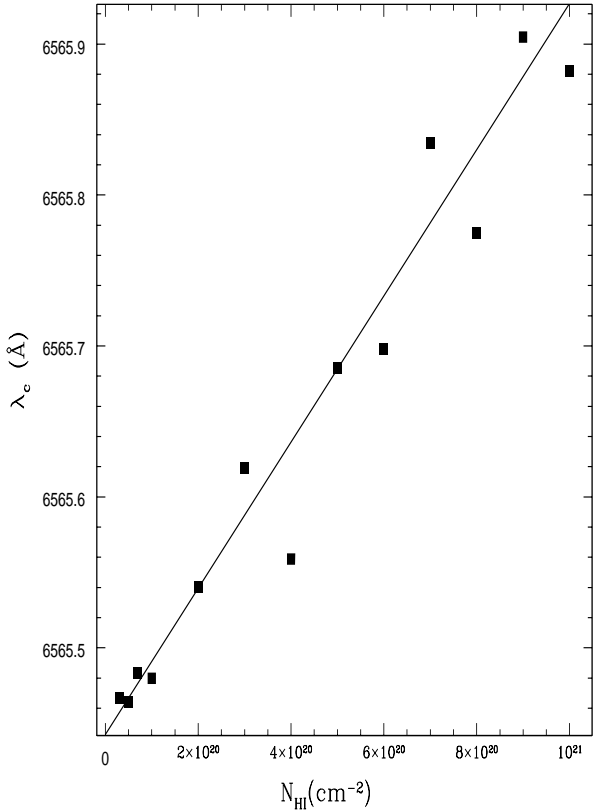, height=11cm, width=8cm}
 \caption{Center locations of H$\alpha$ wings as a function of $N_{HI}$.
The scatter shows the statistical nature of the Monte Carlo calculations.
The center shift amounts only $20{\rm\ km\ s^{-1}}$ for a change
of $N_{HI}$ from $10^{20}{\rm\ cm^{-2}}$ to $10^{21}{\rm\ cm^{-2}}$. 
}
\end{figure}
 In Fig.~4, we plot the center locations as a function of $N_{HI}$.
The result shows the redward wing shift of H$\alpha$ wings as $N_{HI}$
increases.  However, the center shift amounts to only 
$20{\rm\ km\ s^{-1}}$ for a change
of $N_{HI}$ from $10^{20}{\rm\ cm^{-2}}$ to $10^{21}{\rm\ cm^{-2}}$. 
Furthermore, the exact center location is ambiguous as is shown by
the considerable scatter in Fig.~4, which is mainly due to the statistical
nature of our Monte Carlo calculations. From this simulation, we may
conclude that the center shift of H$\alpha$ wings in this range of 
$N_{HI}$ does not have much observational significance.

\subsection{Raman Scattered He~II~6545}

Assuming a similar nebula condition discussed in Section 2,
we prepare He~II~$\lambda$~1025 line as an input spectrum 
represented by a single Gaussian with 
a width of $\Delta\lambda= 0.08{\rm\ \AA}$ shown in Fig.~2. 
As is discussed later, this choice explains fairly well the spectrum
around H$\alpha$ of the symbiotic star V1016 Cyg obtained with the
3.6 m Canada-France-Hawaii Telescope in 2002 (e.g. Lee et al. 2003).
First, we investigate the Raman scattered He~II~6545 feature
without any contribution from the UV continuum. 

\begin{figure}
\epsfig{file=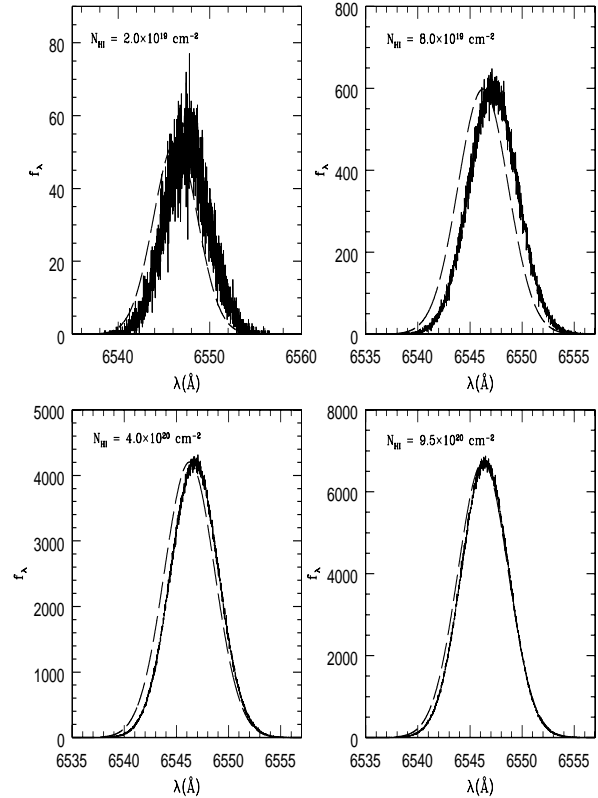, height=11cm, width=8cm}
 \caption{
Profiles of the Raman scattered He~II~$6545$ 
from a scattering region with 
$N_{HI} = 2\times 10^{19}, 8\times 10^{19},4\times 10^{20}$, and
$9.5\times 10^{20}{\rm\ cm^{-2}}$.
The incident He~II~1025 emission line is assumed to be given by a
single Gaussian shown in Fig.~2. 
Long dashed lines show Gaussian line profiles of He~II~6545
that would be obtained for a constant Raman conversion rate.
Note that the peak locations
move redward by different amounts for various $N_{HI}$.
} 
\end{figure}

In Fig.~5, we show the profiles of the Raman scattered He~II~$6545$ 
from a scattering region with neutral hydrogen column densities
$N_{HI} = 2\times 10^{19},8\times 10^{19},4\times 10^{20}$,
and $9.5\times 10^{20}{\rm\ cm^{-2}}$. 
In order to clarify the shift of the peak position, 
by long dashed lines we show the single Gaussian line profile
of the Raman scattered He~II~6545, which would appear in the case
where the Raman conversion rate is constant.  The Monte Carlo results 
are shown by solid lines.  In this figure,
we immediately see that the center location moves redward or toward
H$\alpha$.  The result for $N_{HI}=2\times
10^{19}{\rm\ cm^{-2}}$ is quite noisy due to low photon statistics.

\begin{figure}
\epsfig{file=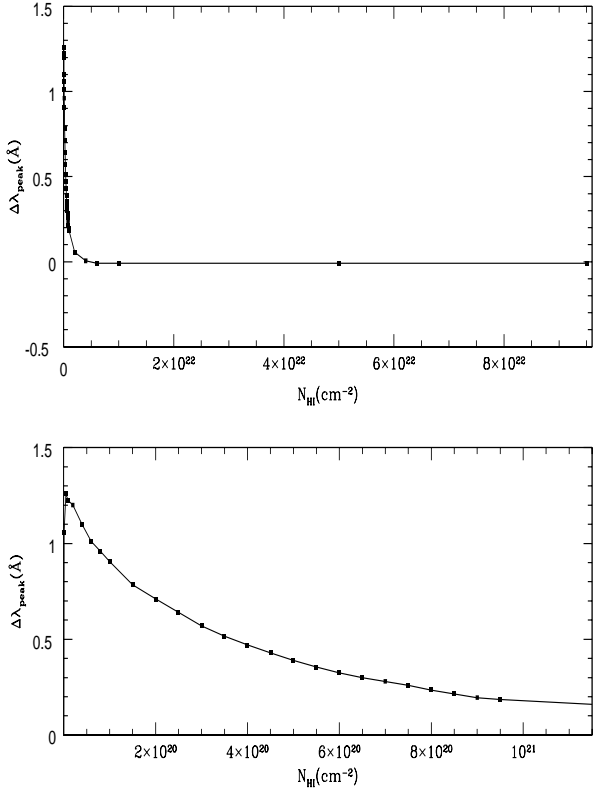, height=11cm, width=8cm}
 \caption{
Center shift of the Raman scattered He~II~6545 feature 
as a function of $N_{HI}$.
The amount of center shift decreases as $N_{HI}$ increases, and
no center shift is seen for $N_{HI}>10^{22}{\rm\ cm^{-2}}$. 
}
\end{figure}

A least square method was applied to the Monte Carlo results in order to
obtain the fits and locate the peak positions of the Raman scattered features. 
We note that all the Raman features are
fitted satisfactorily by a Gaussian function with a width
$\Delta\lambda_{6545}=6.4^2\times 0.08{\rm\ \AA}=3.3
{\rm\ \AA}$.  

In Fig.~6, we  plot the amount of center shift as a function of $N_{HI}$.
For a column density $N_{HI}\sim 10^{20} {\rm\ cm^{-2}}$ 
the peak position shift amounts to $\Delta\lambda_0\sim
0.9{\rm\ \AA}$, which is quite significant and similar to the value
discussed in the previous section. 
The amount of center shift decreases as $N_{HI}$ increases in the region 
$N_{HI}> 5\times 10^{19}{\rm\ cm^{-2}}$, and
no center shift is seen for $N_{HI}>10^{22}{\rm\ cm^{-2}}$. At this high
column density, the Raman conversion rate for He~II~$\lambda$~1025
saturates to a constant value $\sim 0.6$, which explains no shift of the
peak position.  The decrease of the peak shift 
as $N_{HI}$ indicates the decrease of the slope of 
the Raman conversion rate as a function of the scattering optical depth, 
which approaches zero in the optically thick limit.
 
For lower column densities, the peak shifts are quite large 
exceeding $1{\rm\ \AA}$. This implies that in an optically thin limit,
the Raman conversion rate is not given by a simple relation that is 
proportional to $(\lambda-\lambda_{Ly\beta})^{-2}$.
It appears that more redward peak shift in low column 
densities may be explained by introducing a parameter $a$ so that
the Raman conversion rate is described in terms of this parameter by
\begin{equation}
p_1(\lambda)= 
{p_0 e^{-[(\lambda-\lambda_{He^+})/
\Delta\lambda]^2}\over (\lambda-\lambda_{Ly\beta})^{2}-a},
\end{equation}
where the peak location coincides with one of the roots of the derivative
$p_1'(\lambda)$.
The introduction of a new parameter $a$ may be justified by the following
argument. In an optically thin scattering region characterized by the
total scattering optical depth $\tau$ and the Raman branching ratio $r_b$,
the Raman conversion efficiency including multiple scattering effects may be
approximated by an infinite geometric series
\begin{eqnarray}
C_{Ram} &\simeq & \tau r_b +\tau(1-r_b)\tau r_b +[\tau(1-r_b)]^2\tau r_b +\cdots
\nonumber \\
&=&{\tau r_b \over 1-\tau(1-r_b)}.
\end{eqnarray}
Since we approximately have $\tau\propto (\lambda-\lambda_{Ly\beta})^{-2}$, 
this leads to a result in the functional form
$C_{Ram}\propto [(\lambda-\lambda_{Ly\beta})^2-a]^{-1}$.
With the assumption of an optically thin medium, it follows that
$a\ll Y^2$.

By differentiating $p_1(\lambda)$ and defining 
$\delta y\equiv \lambda-\lambda_{Ly\beta}$, we 
obtain a cubic equation in $\delta y$, which is
\begin{equation}
(\delta y)^3+Y(\delta y)^2+ (\Delta\lambda^2-a)(\delta y)-aY=0.
\end{equation}
Upon investigating the roots of the cubic equation for various values of $a$,
we find that when $a\sim 0.09{\rm\ \AA}^2$, the peak shift can
be as large as $1.2{\rm\ \AA}$ found for $N_{HI}=0.95\times10^{19}
{\rm\ cm^{-2}}$ in our Monte Carlo calculation.

In the cases of optically thick scattering regions, the 
Raman conversion rate depends on $N_{HI}$ in a non-linear way,
which gives rise to the various amount of the peak shift
for $N_{HI}$. In turn, the nonlinearity is attributed to the multiple
Rayleigh scattering effect before Raman conversion takes place. 

\begin{figure}
\epsfig{file=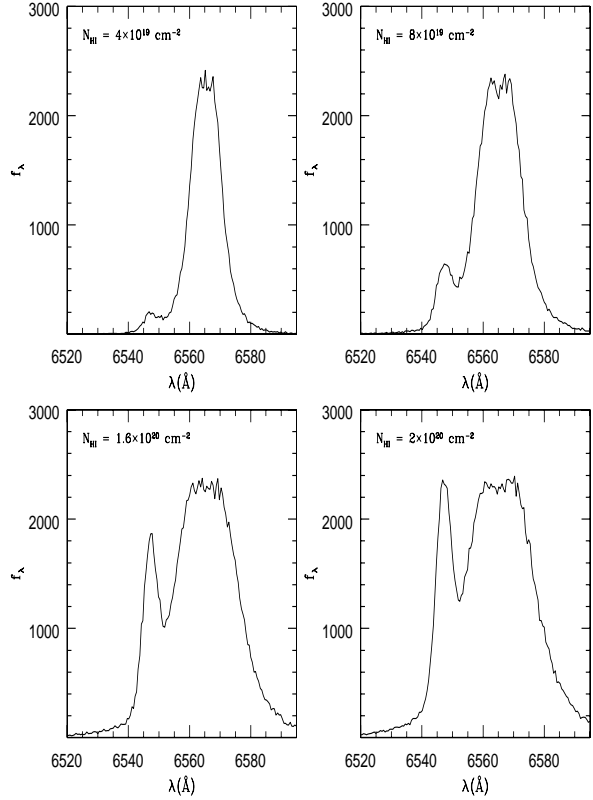, height=11cm, width=8cm}
 \caption{
Combined spectra of UV continuum and He~II~$\lambda$~1025 that are
Raman scattered from scattering regions with various $N_{HI}$.
The strengths of both the H$\alpha$ wings and the He~II~$\lambda$~6545 
feature vary for different values of $N_{HI}$, or the scattering
optical depth.  Due to the unknown covering factor of the scattering
region, these simulated profiles should be multiplied by a constant factor
to fit an observed spectrum.
}
\end{figure}

In Fig.~7, we show combined spectra of UV continuum and He~II~$\lambda$~1025
for various $N_{HI}$.
The strengths of both the H$\alpha$ wings and the He~II~$\lambda$~6545 
feature vary as the column density of the scattering
region mainly due to the difference in scattering cross section. 
However, the observed strength is directly proportional to the covering
factor of the scattering region, and hence these two factors are
degenerate. 
If we fix the covering factor and geometrical shape of the
scattering region, the Raman conversion efficiency
is determined by the scattering optical depth in a non-linear
way.  Therefore, with the unknown covering factor of the scattering
region, these simulated profiles should be multiplied by a constant factor
to fit an observed spectrum.

\begin{figure}
\epsfig{file=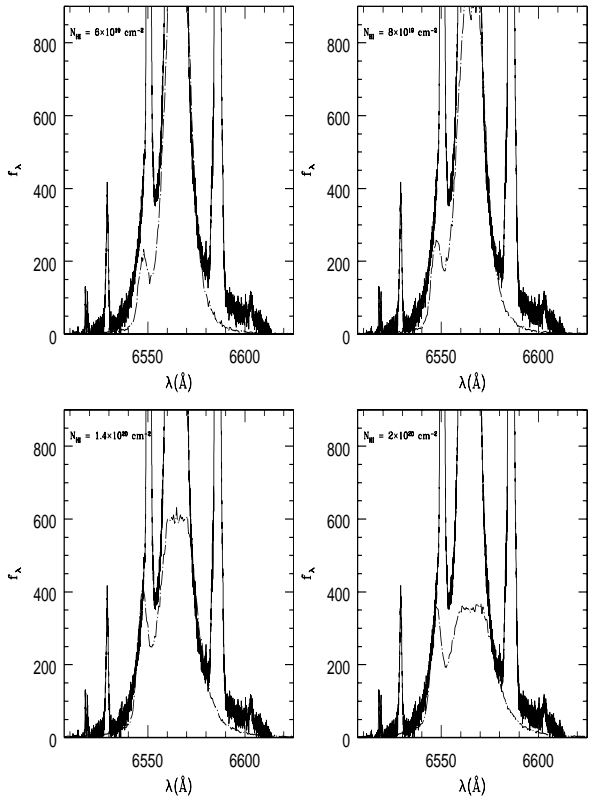, height=11cm, width=8cm}
 \caption{
A spectrum of the symbiotic star V1016~Cyg obtained
with the 3.6~m Canada-France-Hawaii Telescope
The simulated H$\alpha$ wings and He~II~6545 feature are overplotted
represented by dotted lines.
}
\end{figure}

In Fig.~8, we show the spectrum of the symbiotic star V1016~Cyg obtained
with the 3.6~m Canada-France-Hawaii Telescope (see Lee et al. 2003 for
observation details). We overplot the simulated H$\alpha$ wings and 
He~II~$\lambda$~6545 features for various column densities 
onto this spectrum. It appears that the simulated profiles with $N_{HI}$
in the range considered in this figure 
provide good fits to the observed spectrum blueward of
the [N~II]~6548 line.  Therefore, in order to determine $N_{HI}$, it
is necessary to locate the exact peak position, which can be done
only after a very careful de-blending of [N~II]~$\lambda$~6548.

From the fact that He~II~6545 feature is observed outside the region where
the Raman conversion rate is saturated and constant, 
the scattering region should be characterized by $N_{HI} <10^{21}
{\rm\ cm^{-2}}$. Noting that Raman scattered O~VI 6825
and 7082 features are formed in a region with $N_{HI}\ge 10^{22}{\rm\
cm^{-2}}$, the scattering region responsible for He~II~6545 should 
include the O~VI scattering region and be much more extended.

\section{Discussion and Observational Ramifications}

\subsection{Electron Scattering Wings}

\begin{figure}
\epsfig{file=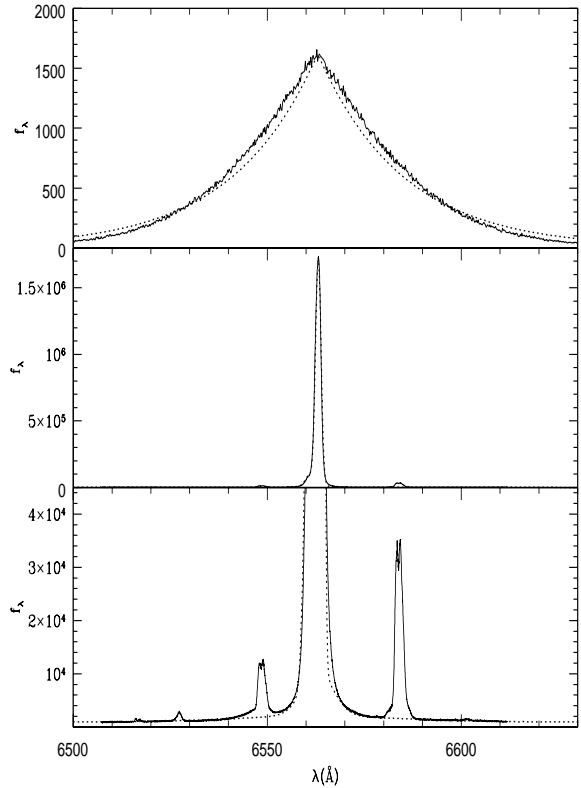, height=11cm, width=8cm}
 \caption{Electron scattering wings overplotted to the spectrum of
   V1016~Cyg obtained with the CFHT.  
The covering factor of the electron scattering region is assumed
to be $C_{el} = 0.05$ and the electron column density $N_e =
10^{24}{\rm\ cm^{-2}}$, for which the Thomson optical depth $\tau_e
= 0.67$. The temperature of the electron cloud is chosen to be
$T_e = 2.5\times 10^4{\rm\ K}$. The top panel shows the electron 
scattering wing represented by a solid line, which was fitted using
the theoretical profile shown by a dotted line investigated by 
Weymann (1970).  }
\end{figure}

Electron scattering of H$\alpha$ line photons is one of physical
mechanisms that have been proposed to be the origin 
of the broad H$\alpha$ wings in many celestial objects including
active galactic nuclei and Wolf-Rayet stars (e.g. Mathis 1970,
Weymann 1970, Hillier 1991).
We generated electron scattering wings using a similar Monte Carlo technique, 
which is shown in Fig.~9.  

Because the H$\alpha$ emission line is 
well-approximated by a single Gaussian $f_\lambda
=1650 e^{-(\lambda-\lambda_{H\alpha})/\Delta\lambda_{H\alpha}}$,
we used this Gaussian as an incident radiation.
The temperature of the electron cloud is chosen to be
$T_e = 2.5\times 10^4{\rm\ K}$.
The covering factor of the electron scattering region is assumed
to be $C_{el} = 3.5\times 10^{-2}$ and the electron column density 
$N_e = 10^{24}{\rm\ cm^{-2}}$, for which the Thomson optical depth $\tau_e
= 0.67$.  It is also noted that similar profiles will be obtained
by a choice of $\tau_e$ and $C_{el}$ with their product fixed
as far as the electron plasma is optically thin.

In the top panel, we show the electron scattering wing produced
by a Monte Carlo technique, which is represented by a thin solid line. 
According to Weymann (1970), electron scattering wings formed in an 
optically thin region may be approximated by exponential profiles
represented by a thick solid line with a functional form
\begin{equation}
f_e(\lambda)=f_0 e^{-(\lambda-\lambda_{H\alpha})/\Delta\lambda_e}.
\end{equation}
Here, the profile width $\Delta\lambda_e=22{\rm\ \AA}$ 
coincides with the Doppler width corresponding to the thermal electron 
velocity with $T_e=2.5\times 10^4{\rm\ K}$,
from which the electron temperature is strongly constrained.
The apparent discrepancy with the theoretical profile is attributed to our 
choice of rather high $\tau_e$
and the fact that the incident H$\alpha$ emission profile
is not a Dirac delta function but a Gaussian with a finite width.

In the middle and bottom panels, the Monte Carlo results are 
represented by dotted lines and the solid lines show the CFHT spectrum
of V1016~Cyg. The fit is quite satisfactory.  
However, the Raman scattering origin of H$\alpha$ wings
naturally explains the existence of the He~II~1025 feature. A more
stringent constraint may be put from careful analyses of the broad H$\beta$
wings and He~II~4850 feature, because Raman scattering cross sections
and branching ratios differ from those corresponding to features
around H$\alpha$. Future spectroscopic observations covering both
H$\alpha$ and H$\beta$ regions simultaneously may shed much light on this
point.

\subsection{Observational Ramifications}

An interesting possibility is that if the He~II emission is not
isotropic but the continuum is, then the equivalent width of 
He~II~$\lambda$~1025
measured by a hypothetical observer located in the neutral scattering
region does not have to be equal to that measured by an observer on Earth.
According to Mastrodemos \& Morris (1998), some fraction of the stellar
wind from the giant component may be captured by the white dwarf
component in symbiotic stars to form an accretion disk. If we accept this
picture, the line emission region may take a disk-like shape,
where more line radiation is expected to the normal direction of the
orbital plane or the accretion disk. Using the ORFEUS data, 
Schmid et al. (1999) pointed out that O~VI~$\lambda\lambda$~1032, 1038
doublets exhibit different profiles from those of their Raman
scattered counterparts, which appears to support this picture.
In this case, for an observer
with a large inclination to the orbital plane, the equivalent width
that is estimated from the Raman scattered He~II 6545 will be smaller
than that estimated from the far UV spectroscopy performed for the
same object.

In the work of Lee et al. (2003), the neutral column density $N_{HI}$
was assumed to be the value that gives a unit scattering optical depth,
of which the validity has not been pursued in a rigorous way. Therefore,
if this choice of $N_{HI}$ is lower than the true value, then the
covering factor of the scattering region should be larger than
they estimated, and vice versa.
However, in order to locate the peak position very accurately,
we have to subtract [N~II]~$\lambda$~6548 feature reliably
using the 3 times stronger [N~II]~$\lambda$~6584, which requires
very careful data reductions. [N~II] lines are weak or absent in 
the higher nebular density S-type systems, for which case the line center
of the Raman scattered He~II$\lambda$~6545 can potentially be determined
without this complication.
As is shown by this study, an accurate value of the
mass loss rate of the giant can be obtained by a more realistic 
model equipped with
a range of $N_{HI}$ with independent information about the equivalent 
width of He~II~$\lambda$~1025. 

It is particularly notable that the continuum photon number flux is 
diluted by a
factor of $(\lambda_i/\lambda_o)^2$ due to the incoherency of the
Raman scattering process. Because the equivalent width is a measure of
the line photon number flux normalized by the continuum photon number
flux, the computation of the equivalent width of the
He~II~$\lambda$~1025 presented by Lee et al. (2003) is erroneous and
overestimated by a factor of 6.4. The correct equivalent
width should be about 0.51 \AA\ which is similar
to the value  adopted in the current work. 

It is very interesting to note that the asymmetric branching ratio of
the Raman scattering cross section gives rise to asymmetric wing profiles
with the stronger red part around H$\alpha$. Regarding this,
spectropolarimetric studies of Raman scattered O~VI 6825 and 7082
features by Harries \& Howarth (1996) show the flip of polarization
direction occurring in the far red part, implying the existence of a
third emission (or scattering) component moving away from the main
scattering region (or emission region). 
Therefore, when examining this kind of component that is moving
away, the effect from the asymmetry of atomic physics should be
corrected first. More spectropolarimetry will be very useful 
to investigate the existence
of similar phenomenon in H$\alpha$ wings.  

Less scattering cross sections and more branching possibilities may
lead to more diverse features in Raman scattered lines around
H$\beta$ and higher Balmer series lines. Therefore, simultaneous
spectroscopy for the Raman scattered features around H$\beta$ and higher
Balmer series lines will shed more light on the mass loss processes
and the ionization structures of symbiotic stars. Similar center shift
phenomena are expected to Raman scattered He~II features formed blueward of
higher Balmer series lines. Because center shifts are also sensitive
to relative kinematics between the UV emission source region
and the scattering region, careful data analyses will be required.
We conclude that Raman features provide very important
and interesting diagnostic tools for investigating the mass loss
processes and ionization structures in symbiotic stars.

\section*{Acknowledgments}
We are very grateful to an anonymous referee for providing helpful
and constructive
suggestiongs that improve the presentation of this paper.
This work is a result of research activities of the Astrophysical 
Research Center for the Structure and Evolution of the Cosmos (ARCSEC) 
funded by the Korea Science and Engineering Foundation.

\label{lastpage}


\begin{thebibliography}{99}

\bibitem[Arrieta \& Torres-Peimbert 2003]{arr03} Arrieta, A. \&
Torres-Peimbert, S., 2003, ApJS, 147, 97
\bibitem[Balick 1989]{bal89} Balick, B., 1989, AJ, 97, 476
\bibitem[Birriel et al. 1998]{bir98} Birriel, J., Espey, B. R., \& 
Schulte-Ladbeck, R. E., 1998, ApJ, 507, 75 
\bibitem[Birriel et al. 2000]{bir00} Birriel, J., Espey, B. R., \& 
Schulte-Ladbeck, R. E., 2000, ApJ, 545, 1020
\bibitem[Espey et al. 1995]{esp95} Espey, B. R., Schulte-Ladbeck, R. E., 
Kriss, G. A., Hamann, F., Schmid, H. M., Johnson, J. J., 1995, ApJ, 454, L61
\bibitem[Harries and Howarth 1996]{har96} Harries, T. J., \& Howarth, I. D.
1996, A\&AS, 119, 61
\bibitem[Hillier 1991]{hil91} Hillier, D. J., 1991, A\&A, 247, 455
\bibitem[Kenyon, 1986]{ken86} Kenyon, S. J., 1986,
The Symbiotic Stars (Cambridge: Cambridge Univ. Press)
\bibitem[Lee 2003]{lee03} Lee, H. -W. 2003, ApJ, 594, 637
\bibitem[Lee 2000]{lee00} Lee, H. -W. 2000,
ApJ, 541, L25
\bibitem[Lee \& Hyung 2000]{leh00} Lee, H.-W., \& Hyung, S., 2000, ApJ,
530, L49
\bibitem[Lee \& Lee 1997]{lee97} Lee, H.-W., \& Lee, K. W. 1997,
MNRAS, 287, 211
\bibitem[Lee et al. 2003]{lee03} Lee, H.-W., Sohn, Y.-J., Kang, Y. W.,
\& Kim, H.-I., 2003, ApJ, 598, 553
\bibitem[Mastrodemos \& Morris 1998]{mas98} Mastrodemos, N. \& Morris, M.,
  1998, ApJ, 497, 303 
\bibitem[Mathis 1970]{mat70} Mathis, J. S., ApJ, 162, 761
\bibitem[Nussbaumer, Schmid, \& Vogel 1989]{nus89}  Nussbaumer, H.,
Schmid, H. M.\& Vogel, M.,1989,A\&A, 221, L27
\bibitem[P\'equignot et al. 1997]{peq97} P\'equignot, D., Baluteau, J.-P.,
Morisset, C., Boisson, C., 1997, A\&A, 323, 217
\bibitem[Seaquist et al. 1984]{sea84} Seaquist, E. R., Taylor, A. R.,
  \& Button, S., 1984, 284, 202
\bibitem[Skopal 2003]{sko03} Skopal, A, 2003,
Recent. Res. Devel. Astronomy \& Astrophys., 1, 111-135, 
Research Signpost, Kerela
\bibitem[Schmid 1989]{sch89} Schmid, H. M. 1989,A\&A,  211, L31
\bibitem[Schmid 1995]{sch95} Schmid, H. M. 1995, MNRAS,  275, 227 
\bibitem[Schmid 1996]{sch96} Schmid, H. M. 1996, MNRAS,  282, 511
\bibitem[Schmid \& Schild 1994]{sch94} Schmid, H. M. \& Schild, H., 1994, 
A\&A,  281, 145
\bibitem[Schmid et al. 1999]{sch99} Schmid, H. M. et al.,  1999, A\&A, 348, 950
\bibitem[Schmid et al. 2000]{sch00} Schmid, H. M., Corradi, R., Krautter, J.,
Schild, H., 2000, A\&A, 355, 261
\bibitem[Schwarz \& Corradi 1992]{sch92} Schwarz, H. E. \& Corradi, R. L. M.,
1992, A\&A, 265, L37
\bibitem[Schwarz \& Monteiro 2003]{sch03} Schwarz, H. E. \& Monteiro, H.,
2003, Rev. Mex. Astron. Astrofis., 15, 23
\bibitem[Taylor \& Seaquist 1984]{tay84} Taylor, A. R. \& Seaquist, E. R., 
1984, ApJ, 286, 263
\bibitem[Van de Steene et al. 2000]{vds00} Van de Steene, G. C.,
Wood, P. R., \& van Hoof, P. A. M. 2000, in Asymmetrical Planetary
Nebulae II: From Origins to Microstructures, ed. J. H. Kastner,
N. Soker, \& S. Rappaport (San
Francisco: ASP Conference Series, Vol. 199), p. 191
\bibitem[van Groningen 1993]{gro93} Van Groningen, E., 1993, MNRAS, 264, 975
\bibitem[Weyman 1970]{wey70} Weymann, R., 1970, ApJ, 160, 31
\bibitem[Yoo, Bak, \& Lee 2002]{yoo02} Yoo, J. J., Bak, J.-Y., \& Lee
  H. -W., 2002, MNRAS, 336, 467
\end{thebibliography}
\end{document}